\documentstyle[prl,tighten,aps]{revtex}

\def\beqa{\begin{eqnarray}}
\def\eeqa{\end{eqnarray}}
\def\beq{\begin{equation}}
\def\eeq{\end{equation}}

\begin{document}

     \title{ Non-Newtonian Gravity, Fluctuative Hypothesis and the
     Sizes of Astrophysical Structures}

\author{Salvatore Capozziello\thanks{
E-mail:capozziello@sa.infn.it}, Salvatore De
Martino\thanks{E-mail:demartino@sa.infn.it}, Silvio De
Siena\thanks{E-mail:desiena@sa.infn.it} and Fabrizio
Illuminati\thanks{E-mail: illuminati@sa.infn.it}
\\ {\em Dipartimento di Fisica ``E. R.
Caianiello",}\\  {\em INFN, Sez. di Napoli and INFM Unit\'a di
Salerno, \\ Universit\`a di Salerno, I-84081 Baronissi (SA)
Italy,}}

\date{\today}
\maketitle

          \begin{abstract}
We show that the characteristic sizes of astrophysical and
cosmological structures, where gravity is the only overall
relevant interaction assembling the system,  have a
phenomenological   relation to the  microscopic scales whose order
of magnitude is essentially ruled by the Compton wavelength of the
proton. This result agrees with the absence of screening
mechanisms for the gravitational interaction and could be
connected to the presence of Yukawa correcting terms in the
Newtonian potential which introduce typical interaction lengths.
Furthermore, we are able to justify, in a straightforward way, the
Sanders--postulated mass of a vector boson  considered in order to
obtain  the characteristic sizes of galaxies.

           \end{abstract}

\vspace{20. mm} PACS: 03.65.Bz;98.70.Vc;98.80-k;98.80.Hw

\section{\normalsize \bf Introduction}

Explaining the  large scale structures of the Universe is one of
the hardest task of modern astrophysics since the growing amount
of observations seems to escape any coherent scheme able to
connect all the parts of the puzzle.

Essentially, from the fundamental physics point of view, we would
like to re-conduct cosmic structures and their evolution to some
unifying theory in which all the today observed interactions are
treated under the same standard. In this case, what we observe
today on cosmological and
 astrophysical scales would be just a
consequence of quantum fluctuations at early epochs. Then, we
should seek for some ``enlarging'' mechanism which, after one (or
more than one) symmetry breaking would be capable of yielding
structures like clusters of galaxies, galaxies and then stars from
primordial quantum spectra of perturbations.

The so called ``inflationary paradigm'' \cite{kolb} related to
several unifying theories (e.g. superstrings, GUT, SUSY, and so
on) should be successful if some ``experimentum crucis'' would
select the right model.

On the other hand, particle physicists need cosmological
predictions and observations since the energies for testing
unified theories are so high that it is extremely unlikely they
will be ever reached on Earth--based laboratories.

As a matter of fact, cosmology needs particle physics and vice
versa. The point is that remnants of primordial epochs should be
found by cosmological observations and, by them, one should
constrain elementary particle physics models.

This philosophy has been pursued by several researchers; first of
all by Sakharov \cite{sakharov} who in 1965 argued that quantum
primordial fluctuations should have expanded towards the present
epoch leading first to classical energy--density perturbations
and, after the decoupling from the cosmological background, to the
observed galaxies, clusters, super-clusters of galaxies, and
afterwards stars. Shortly, the underlying issue of any modern
theory of cosmological perturbation is this: primordial quantum
fluctuations should be enlarged by cosmological dynamics to the
present large scale structures. Now the problem is not only
whether observations agree with this scheme (e.g. COBE and IRAS
data or large scale structure surveys \cite{hucra}) but, mainly,
whether the astrophysical and cosmological systems ``remember''
their quantum origin or not.

Despite of the apparent sharp division of the classical and
quantum worlds, macroscopic quantum phenomena exist and some
behaviors of classical systems can be explained only in the
framework of quantum mechanics. The high $T_{c}$
superconductivity and several other macroscopic coherent systems
(e.g. optical fibres) are famous instances of these peculiar
phenomena in which a quantum ``memory'' persists at the
macroscopic scale.

Recently, a new intriguing conjecture has been proposed to find
signatures of $h$ at the classical, macroscopic scales: in
\cite{calogero}, it has been argued about a possible gravitational
origin of quantization emerging thanks to the universal
interaction of every particle in the Universe with the
gravitational stochastic background field generated by all other
particles.

In the framework of this fluctuative Machian scheme, it is
possible to show how classical nonlinearity and chaoticity of the
gravitational interaction  yields a characteristic unit of action
per particle that coincides, in order of magnitude, with $h$.

Further studies \cite{demartino} have generalized this scheme to
the other fundamental interactions responsible for  macroscopic
structures, finding that $h$ is the characteristic action per
particle also for macroscopic systems not bound by gravitational
interactions but by other forces, e.g. electromagnetic.

In this scheme, classical laws of force  describing the
interactions among the constituents of $N$--particle systems of
mean length scale $R$, lead to a quantum  characteristic action
per particle. The forces considered can be, for instance, the
electromagnetic interactions between charged particles in large
macroscopic systems as charged beams in particle accelerators,
plasmas, and neutral dipolar crystals, or the strong interactions
between quarks in hadronic bound aggregates, and so on.

Having such a procedure at hand, it seems very natural  to
investigate whether it can be applied to determine the existence
of quantum  signatures or ``memories'' for astrophysical
structures.

In a recent paper by the authors \cite{cap}, it was shown that the
order of magnitude of characteristic observed radii of typical
galaxies can be inferred starting from the microscopic
fundamental scales through a phenomenological scaling law.
Further scaling relations can be derived, showing that such a
connection can equivalently be obtained either by considering the
nucleons as the elementary constituents of a galaxy, or, as usual
in astrophysics, by considering stars as its granular components.
The key issue is that we can define an interaction length
$\lambda$ and a typical number of elementary constituents $N$
(e.g. $N_n$ the number of nucleons with $\lambda_c$, the Compton
wavelength,  or $N_s$ the number of stars with $\lambda_s$, the
typical interaction length of a star, for example the ''size" of
the Solar System) and, by the relation

\beq\label{compton} R\simeq\lambda\sqrt{N}\,,\eeq to obtain the
observed typical size (in the case of galaxies we obtain $\sim
10$ kpc a typical galactic scale).

In the present paper, we show that such a relation holds for other
self-gravitating systems as stellar globular clusters, clusters
and super-clusters of galaxies and we show that it can be
implemented by taking into account a gravitational interaction
explicitly scale-dependent, as predicted by several effective
quantum field theories in the low energy limit.

More precisely, several renormalizable quantum theories of gravity
require a modification, in the low energy limit, of Newton
potential. Furthermore, if we do not require enormous amounts of
dark matter as the only mechanism to explain the puzzle of the
present day astrophysical observations, a scale--dependent
gravitational interaction is  needed. This point of view is
supported by several authors \cite{mcgaugh} which state that
recent experiments on cosmic microwave background as BOOMERANG
\cite{debernardis} or astrophysical structures can be explained
by taking into account modified Newtonian dynamics
\cite{sanders},\cite{eckhardt}.

The paper is organized as follows. In Sec. 2, we deal with the
weak field limit of extended theories of gravity which, in
general, gives rise to Yukawa-like corrections to the Newtonian
potential, which, naturally imply characteristic lengths. Sec. 3
is devoted to the discussion of the characteristic sizes of
astrophysical structures. In Sec. 4, we discuss how
time-statistical fluctuations of granular components of
self-gravitating systems are related to their characteristic
sizes. Conclusions are drawn in Sec.5.

\section{\normalsize \bf   Newtonian limit of extended theories of gravity
and the emergence of characteristic interaction lengths}

Extended theories of gravity have become a sort of paradigm in the
study of gravitational interaction since several motivations push
for enlarging the traditional scheme of Einstein general
relativity. Essentially, they consist to add corrections due to
scalar fields or curvature invariants of the form, $\phi^2R$,
$R^2$, $R_{\mu\nu}R^{\mu\nu}$, or $R\Box R$ in the
Einstein-Hilbert gravitational action. Such issues come,
essentially, from cosmology and quantum field theory.

In the first case, it is well known that higher--derivative
theories \cite{starobinsky} and scalar--tensor theories \cite{la}
furnish inflationary cosmological solutions capable, in principle,
of solving the shortcomings of standard cosmological model
\cite{weinberg}.

In the second case, every unification scheme as superstrings,
supergravity or grand unified theories, gives effective actions
where nonminimal couplings to the geometry or higher order terms
in the curvature invariants come out. Such contributions are due
to one--loop or higher--loop corrections in the high curvature
regimes near the full (not yet available) quantum gravity regime
\cite{odintsov}. However, in the weak--limit approximation, all
these theories should be expected to reproduce the Einstein
general relativity which, in any case, is experimentally tested
only in this limit \cite{will}.

This fact is matter of debate since several relativistic theories
{\it do not} reproduce Einstein results at the Newtonian
approximation but, in some sense, generalize them. In fact, as it
was first noticed by Stelle \cite{stelle}, a $R^2$--theory gives
rise to Yukawa--like corrections to the Newtonian potential which
could have interesting physical consequences.

For example, some authors  have shown that a conformal theory of
gravity is nothing else but a fourth--order theory containing
such terms in the Newtonian limit and, by invoking these results,
it could be possible to explain the missing matter problem
''without" dark matter \cite{mannheim}.

 In general, it can be shown
\cite{stelle},\cite{schmidt},\cite{kenmoku} that most of the
extended theory of gravity have a weak field limit of the form
\beq \label{yukawa}
V(r)=-\frac{G_{\infty}m}{r}\left[1+\sum_{k=1}^{n}\alpha_k
e^{-r/\lambda_k}\right]\,, \eeq where $G_{\infty}$ is the value of
the gravitational constant as measured at infinity, $\lambda_k$ is
the interaction length of the $k$-th component of non-Newtonian
corrections. The amplitude $\alpha_k$ of each component is
normalized to the standard Newtonian term (see
\cite{will},\cite{principia} for further details). The discussion
involves also the variation of the gravitational coupling. As an
example, let us take into account only the first term of the
series in $(\ref{yukawa})$ which is usually considered the leading
term (this choice is not sufficient if other corrections are
needed). We have

 \beq
\label{yukawa1} V(r)=-\frac{G_{\infty}m}{r}\left[1+\alpha_1
e^{-r/\lambda_1}\right]\,. \eeq The effect of non-Newtonian term
can be parameterized by $(\alpha_1\lambda_1)$. For large
distances, at which $r\gg\lambda_1$, the exponential term
vanishes and the gravitational coupling is $G_{\infty}$. If
$r\ll\lambda_1$, the exponential becomes unity and, by
differentiating, we get

\beq \label{yukawa2}
G_{lab}=G_{\infty}\left[1+\alpha_1\left(1+\frac{r}{\lambda_1}\right)e^{-r/\lambda_1}\right]
\simeq G_{\infty}(1+\alpha_1)\,, \eeq where
$G_{lab}=6.67\times10^{-8}$ g$^{-1}$cm$^3$s$^{-2}$ is the usual
Newton constant measured by Cavendish-like experiments. Of
course,  $G_{\infty}$ and $G_{lab}$ coincide in standard gravity.
It is worthwhile to note that, asymptotically, the inverse square
law holds but the measured coupling constant differs by a factor
$(1+\alpha_1)$. In general, any exponential correction introduces
a characteristic length that acts at a certain scale for the
self-gravitating systems.

This approach has been pursued by several authors who  tested
non-Newtonian corrections by ground-based experiments using
totally different techniques
\cite{fischbach},\cite{speake},\cite{eckhardt1}. The general
outcome of these experiments, even retaining only the term $k=1$,
is that a ''geophysical window" between the laboratory and the
astronomical results has to be taken into account. In fact, the
range

\beq |\alpha_1|\sim 10^{-2}\,,\qquad \lambda_1\sim 10^2\div
10^3\,\mbox{m}\,,\eeq is not excluded at all. The sign of
$\alpha_1$ tells us if corrections are attractive or repulsive.
Another interesting suggestion has been given by Fujii
\cite{fujii1}, which proposed that the exponential deviation from
the Newtonian standard potential (the ''fifth force") could arise
from the microscopic interaction which couples to nuclear isospin
and baryon number.

The astrophysical counterpart of these non-Newtonian corrections
seemed ruled out till some years ago due to the fact that
experimental tests of general relativity predict ''exactly" the
Newtonian potential in the weak energy limit, ''inside" the Solar
System. Recently, as we said above,
 indications of an anomalous, long--range acceleration
revealed from the data analysis of Pioneer 10/11, Galileo, and
Ulysses spacecrafts makes these Yukawa--like corrections come
into play \cite{anderson}. Besides, Sanders \cite{sanders}
reproduced the flat rotation curves of spiral galaxies by using

\beq \alpha_1=-0.92\,,\qquad \lambda_1\sim 40\,\mbox{kpc}\,.\eeq
His main hypothesis is that the additional gravitational
interaction is carried by an ultra-soft vector boson whose range
of mass is $m_0\sim 10^{-27}\div 10^{-28}$eV. The action of this
boson becomes efficient at galactic scales without the request of
enormous amounts of dark matter to stabilize the systems.

On the other hand, by asking for a characteristic length emerging
from the standard theory of cosmological perturbation, it is
possible to explain the observed segregation of hot stellar
systems in the so called {\it fundamental plane} of galaxies
(''ordinary" and ''bright" galaxies) \cite{capaccioli}. In that
case, the length is the ''Jeans length" of the protogalaxy
($\lambda\sim 3\div 10$ kpc) and, due to this characteristic
size,  a Yukawa correction was found in the gravitational
potential with a characteristic interaction ''length" of the same
order of magnitude of that proposed by Sanders.

In this paper, we discuss the emergence of characteristic lengths
by a time-statistical fluctuation of the granular components of
self-gravitating systems. Our guess is that such lengths give rise
to non-Newtonian corrections in the gravitational potential.

In the next section we discuss what we intend for the
characteristic size of a self-gravitating system and then we
discuss the fluctuative hypothesis.

\section{\normalsize \bf Characteristic sizes of
astrophysical self-gravitating systems}

 In general, the concept of ''size" of a self-gravitating system
is not well-based since, in several cases, the boundary cannot be
univocally defined. Let us briefly define globular clusters,
galaxies, clusters and super-clusters of galaxies
 by their typical lengths and masses\footnote{In this paper, we
are taking into account only systems where gravity is the only
overall interaction acting between the components. In this sense,
a star is not a purely self-gravitating system since, inside it,
gravity is balanced by the pressure due to electromagnetic and
nuclear interactions. However, we can take into account stars as
granular constituents of globular clusters and galaxies and
define a typical interaction length as the ''size"  of a
planetary system around a star.}.

\vspace{2. mm}

A {\it globular cluster} is a very compact self-gravitating
stellar system whose typical radius is $R_{gc}\sim 10$ pc . It
contains up to $10^6$ stars ($M_{gc}\sim 10^6 M_{\odot}$) and is
assumed completely virialized due to collisional interactions
between stars.

\vspace{2. mm}

A {\it galaxy} is a collisionless, diffuse gravitating system
without an effective boundary. Astronomers define operative
characteristic sizes as the {\it effective radius} $R_e$ which is
the radius of the isophote containing half of the total
luminosity, or the {\it tidal radius} $R_t$ corresponding to the
distance from the center where the density drops to zero
\cite{binney},\cite{vorontsov}. Other definitions are possible by
using photometry or kinematics but, assuming as a typical
interaction size a length $R_g\sim 1\div 10$ kpc is quite
reasonable from dwarf to giant galaxies \footnote{Several authors
assumes that the halo of giant galaxies can extend as far as
$100$ kpc from the center taking into account the dark matter
component. Here we do not assume any dark matter hypothesis and
do not want to enter into details of galactic dynamics and
morphology. For example, Milky Way, a typical spiral galaxy, has
an observed  disk scale length $R_d\simeq 3.5\pm 0.5$ kpc while
kinematics of globular clusters and $21$ cm-radio observations of
neutral hydrogen give a maximal halo extension of $20\div 30$
kpc. For our purposes, assuming $10$ kpc as a characteristic
size, with a possible error of an order of magnitude, is a good
number.}. Typical masses are $M_g\sim 10^{10\div 12}M_{\odot}$ for
giant galaxies and $M_g\sim 10^{8\div 9}M_{\odot}$ for dwarf
galaxies.

\vspace{2. mm}

As a {\it cluster of galaxies}, following Abell \cite{abell}, we
define a self-gravitating system whose granular components are
galaxies with a typical radius $R_{cg}=R_a\simeq 1.5 h^{-1}$ Mpc
(the Abell radius) and a typical mass $M_{cg}\sim 10^{15}
h^{-1}M_{\odot}$ for rich clusters, where $h$ is the dimensionless
Hubble constant whose value is in the range $0.5<h<1$
\cite{peebles}.

\vspace{2. mm}

A {\it super-cluster} is a self-gravitating system of clusters of
galaxies whose typical size is $R_{sc}\sim 10\div 100 h^{-1}$ Mpc
and typical mass is $M_{sc}\sim 10^{15\div 17} h^{-1}M_{\odot}$.

\vspace{2. mm}

 {\it Groups of galaxies} are systems containing
$10\div 20$ galaxies, as our Local Group, but there are no
evidences that they could be considered self-gravitating systems
and, in any case, they are always part of  more extended cluster
of galaxies (in the case of Local Group, it is a part of the
Virgo Cluster).

\vspace{2. mm}

 The main difference between a
globular cluster and the other  systems is that the former is a
collisional system while the others are collisionless. This fact
implies a completely different dynamical treatment \cite{binney}.

 The properties of these
self-gravitating systems can be deduced by assuming them to be
relaxed and virialized systems where gravity is the only overall
interaction \cite{binney}. This assumption
 is, some times, not completely justified. In
fact, we have to keep in mind that these systems undergo
environmental effects, being never completely isolated; they
always belong to larger gravitationally bound systems and the
observational times are so short that the overall dynamics can
only be extrapolated \cite{binney}, \cite{vorontsov}.

Furthermore, as we said, the dynamics of astrophysical systems
have to be related to  cosmological evolution  so that, in today
observed dynamics, some quantum signature  of  primordial quantum
perturbations should be present \cite{sakharov}.

However the main difficulty, is to provide a physical route
connecting the sizes of astrophysical structures with the
extremely small numbers of quantum mechanics.

As a first step, we can  build a model of the above structures
composed of self--gravitating microscopic constituents (nucleons)
 which undergo some statistical fluctuations \cite{calogero}.

In a semi--quantitative analysis of the model, we introduce as the
only observational input the number of nucleons contained.
 The
characteristic dimension of such a model, as we shall see below,
is a functions of the microscopic nucleon scales (the Compton
wavelength of a nucleon  $\lambda_c\sim 10^{-13}$ cm) and of the
number of microscopic constituents.

The result is that  the characteristic radii so deduced,
numerically, coincide with those observed considering the usual
gravitational constituents  (stars for globular clusters and
galaxies or galaxies themselves for clusters and super-clusters
of galaxies, not nucleons).

Besides,
 we obtain a  scaling relation between
the  units of length and action of the  granular gravitational
components, ranging from nucleons up to stars and galaxies.

On the other hand, the characteristic dimensions of astrophysical
structures appear to be independent of the scale of the
constituents considered. It is only needed that they depend on a
minimal scale of length, which is, in order of magnitude,  the
Compton wavelength of a nucleon.

All these results could suggest a sort of macroscopic quantum
coherence for large scale gravitational systems.

Furthermore, the emergence of these characteristic scales could
have a dynamical counterpart in the non-Newtonian, Yukawa
corrections of gravitational potential as in Eq.(\ref{yukawa}).

\section{\normalsize \bf The fluctuative hypothesis for
self-gravitating systems}

In order to get a general rule to define ''sizes" for
self-gravitating systems, we start by considering the total action
for a bound system  with a very large number $N$ of constituents.
Let $E$ be the total energy. Let ${\cal T}$ be the characteristic
global time of the system (e.g. the time in which a particle
crosses the system, or the time in which the system evolves and
becomes relaxed or virialized). By these two quantities, we get

 \beq
\label{1} A \cong E{\cal T} \, , \eeq which is the total action
of the system. The only hypothesis which we need is that the
system may undergo a time--statistical fluctuation, so that the
characteristic time $\tau$ for the stochastic  motion per
particle is \cite{calogero}, \cite{demartino}

\beq \label{2} {\tau}\cong\frac{{\cal T}}{\sqrt{N}} \, . \eeq
This {\it fluctuative hypothesis} naturally emerges from the fact
that, if $N$ is large, dynamics is affected by some kind of
statistical (chaotic) fluctuation
\cite{calogero},\cite{binney},\cite{camacho},\cite{roy}.

We can then define an energy per granular component

\beq \label{3} \epsilon\cong\frac{E}{N} \, , \eeq so that the
characteristic (minimal) unit of action $\alpha = \epsilon \tau$
per granular component is expressed by the  scaling relation
\cite{calogero}

\beq \label{4} \alpha=\epsilon\tau\cong\frac{A}{N^{3/2}} \, . \eeq
Let us now consider the observational data for the above
self-gravitating systems, in order of magnitude. Since we are
taking into account  virialized systems, we can assume

\beq \label{vir} 2E_{k}+U=0\,, \eeq
 where $E_{k}$ is the kinetic energy and $U$ the gravitational
energy.

The total energy $E$ can then be assumed to be

\beq\label{total} E\simeq E_{k}\simeq N M v^2 \eeq where $N$ is
the typical number of granular components (e.g. stars in a galaxy
or in a globular cluster, or galaxies in a cluster or
super-cluster of galaxies;  $M$ is the typical mass (e.g.
$1M_{\odot}$ for a Main Sequence star or $10^{10\div 11}M_{\odot}$
for a galaxy like Milky Way);
 $v$ is a characteristic typical velocity
which we choose to be the circular speed of the stars in the disk
of galaxies $(\simeq 10^{7\div 8}$cm/sec)  or the velocity
dispersion of the galaxies inside a cluster $(\simeq 10^{8\div
9}$cm/sec). All these numbers are quite accurately measured by
the methods of stellar kinematics, statistics and photometry
\cite{binney}\footnote{Due to the virial theorem and the
conservation of energy, we are assuming that we are dealing with
 almost  isolated systems also if they belong to larger
gravitationally bound systems. We  discard considerations on the
potential energy $U$ which imply statements on  rotation curves,
the dependence of the mass from the radius, and the introduction
of dark matter.}.
 All the above quantities entering into the definitions of the
energy, time and action scales
 are   quantities coming from observations. In particular, nowhere we
introduce the characteristic radius $R$ of the structures since
this is what we wish to predict in the framework of our
considerations.

Let us start by taking into account a galaxy.
 The energy per unit of mass is of the order
$10^{15}$ (cm/sec$)^2$, while the period of a galactic rotation,
which can be assumed as the characteristic global time, is of the
order \cite{binney}

\beq \label{5} {\cal{T}}_{rot} \cong 10^{15}\mbox{sec} \, , \eeq
and finally the total mass of a typical galaxy is of the order
\cite{binney}

\beq \label{6} M_g=N_{s}^{(g)}M_s \cong 10^{44}\mbox{gr} \, . \eeq
From Eq.(\ref{1}), combining these numbers, we get the typical
action

\beq \label{7} A \cong 10^{74}\mbox{erg sec} \, . \eeq The typical
number of nucleons in a galaxy is \cite{binney}

\beq \label{8} N \cong 10^{68} \, . \eeq Inserting these numbers
in Eq.(\ref{4}) we see that, up to an order of magnitude, the
characteristic unit of action $\alpha$ of a  galaxy, considered
as an aggregate of nucleons, is of the order of the Planck action
constant, $h \sim 10^{-27}$ erg sec. It is worthwhile to stress
that also if dark matter is considered, the result does not change
dramatically since the mass to luminosity ratio is of the order
$10 \div 100$.

As a further step, we note that Eq.(\ref{4}), together with the
numerical result $\alpha \cong h$, can be re-formulated as a
scaling relation for the {\it mean} action per microscopic
component $a \equiv A/N$, that is

\beq \label{9} a \cong h \sqrt{N} \, . \eeq We can then deduce
that the fluctuative factor $\sqrt{N}$ provides the rescaling
coefficient from the microscopic scales to the characteristic
macroscopic dimensions.

Let us now take into account the lengths. Given the nucleons as
the basic microscopic constituents in our model, the natural
quantum unit of length associated to each single constituent is
the Compton wavelength $\lambda_c = h/mc$, with $c$ the velocity
of light, and $m \cong m_{p} \cong 10^{-24}$ gr, the proton mass.
 In analogy with
Eq.(\ref{9}), we have, in general,

\beq \label{10} R \cong \lambda_{c} \sqrt{N} \, . \eeq For a
galaxy, with $N$ given by Eq.(\ref{8}),  we obtain

\beq \label{11} R_{g} \cong 10^{21} \div 10^{22} \mbox{cm} \simeq
1 \div 10 \mbox{kpc} \,  \eeq which as we said above is a length
of a galaxy. In particular, the numerical agreement of
Eq.(\ref{10}) with the observed galactic radii, is interesting,
independently of the present derivation, since it links the scale
of a large structure like a galaxy to the Compton wavelength of
the elementary constituents (the nucleons) and to the total
number of such constituents.

It is worth noticing that, for typical galaxies, $R_{g}$ is the
characteristic dimension where their rotation curve can be assumed
flat \cite{binney} and where the halo and the disk stabilize each
other.

The validity of Eq.(\ref{10}) is not restricted to the galaxies,
but provides the correct order of magnitude of the observed radii
also if one considers the other structures which we mentioned
above.

In the case of globular clusters, considering the right $N$
(which we easily deduce by the number of stars which constitute
them, i.e. $10^6$), we get

\beq \label{11a} R_{gc} \cong 10^{18} \div 10^{19} \mbox{cm}
\simeq 1 \div 10 \mbox{pc} \, . \eeq

For clusters of galaxies we obtain $R_{cg}\cong 1$ Mpc and for
super-clusters $R_{sg}\cong 10\div 100$ Mpc.

The discussion can be extended to the whole Universe, and to other
astrophysical objects, such as planetary systems  (like the Solar
System), provided one inserts in Eq.(\ref{10}) the correct value
of the number of nucleons $N$ contained in such structures.

These findings indicate that the quantum parameter $\lambda_{c}$
and  the number of nucleonic constituents $N$, determine the
observed astrophysical and cosmological dimensions.

 The crucial objection to these results would be
that stars or single galaxies, rather than nucleons, are the
natural candidates as elementary gravitational constituents of a
typical galaxy or a typical cluster of galaxies.

This apparent difficulty can be solved deriving a simple scaling
law, which holds true at any scale. Let us take into account, for
example, the number of stars $N_{s}^{(g)}$ contained in a typical
galaxy, and the number $N_{n}^{(s)}$ of nucleons in a star. We
can then obviously write, for the total number of nucleons in a
typical galaxy

\beq \label{15} N \cong N_{s}^{(g)}N_{n}^{(s)} \, . \eeq By
Eq.(\ref{15}), we can write Eq.(\ref{10}) as

\beq \label{16} R_g \cong \lambda_{s} \sqrt{N_{s}^{(g)}} \, , \eeq
where

\beq \label{17} \lambda_{s}  \equiv \frac{A_{s}}{M_{s}c}\,,\qquad
A_{s} \equiv  h [N_{n}{^{(s)}}]^{3/2} \,,\qquad M_{s} \equiv m
N_{n}^{(s)} \, . \eeq Here, as above, $M_{s}$ is  the total mass
of a star, while the quantity $A_{s}$ is the characteristic unit
of action of a star in the framework of our model, taking the
stars as the elementary constituents of a typical galaxy.
Inserting the numerical values \cite{binney} $N_{n}^{(s)} \cong
10^{57}$, $N_{s}^{(g)} \cong 10^{10} \div 10^{12}$, we obtain

\beq \label{18} \lambda_{s} \cong 10^{13} \div 10^{15} \mbox{cm}
\, , \eeq
 which agrees with the typical range of interaction of a star
(e.g. that of the Solar System), while for $R_g$ we obviously
obtain again the value (\ref{11}).

Therefore, Eqs.(\ref{10}) and (\ref{16}) show that we can derive
the observed galactic radius $R_g$ either by considering a galaxy
as a gas of $N$ nucleons with the fluctuation (\ref{2}) defined
with respect to $N$, or by considering, as usual, a typical
galaxy as a gas of stars and assuming the fluctuative ansatz
(\ref{2}) rescaled with respect to the number of stars
$N_{s}^{(g)}$.

The reason for the validity of this relation  (which, in
principle, holds on any scale)  relies on the existence of a
minimal scale of action which is needed for mechanical stability.
In fact, the numerical value of the unit of action $A_{s}$
defined in Eq.(\ref{17}) is $\cong 10^{58}$ erg s and thus
coincides, in order of magnitude, with the total action for a
typical star. Thus the rescaling relations (\ref{4}) and
(\ref{10}) hold true also for a star, and $\lambda_{s}$ appears
as the effective macroscopic ``Compton wavelength'' of a star.
However $\lambda_s$ is the typical range of interaction also in
the case of a globular cluster giving
$R_{gc}\cong\lambda_{s}\sqrt{N_{s}^{(gc)}}\sim 1\div 10$ pc.

Immediately we derive $\lambda_g$ as the range of interaction for
galaxies considered as granular constituents of clusters and
super-clusters. Analogously, we have

\beq \label{17a} \lambda_{g}  \equiv \frac{A_{g}}{M_{g}c}\,,\qquad
A_{g} \equiv  h [N_{n}{^{(g)}}]^{3/2} \,,\qquad M_{g} \equiv m
N_{n}^{(g)} \, . \eeq and then $\lambda_g\cong 10\div 100$ kpc.
In this case, we can hierarchically consider a cluster or a
super-cluster of galaxies as a gas of nucleons, stars or galaxies.
It is interesting to note that $\lambda_g$ is the observed typical
separation length between galaxies in a cluster.

 At this point, it is straightforward the  connection
to the non-Newtonian gravitational potential (\ref{yukawa}). As
we discussed above, the Yukawa corrections have to be reconducted
to the emergence of typical scales for self-gravitating systems.
For example, as discussed in \cite{sanders} and in
\cite{eckhardt},
  by the interaction ranges of some vector bosons, it is possible
  to explain the flat rotation curves of spiral galaxies without
  asking for large amounts of dark matter. Yukawa corrections naturally
  emerge in relation to these interaction ranges. The main shortcoming of
  their approach is that, till now, no ultra-light vector boson has
  been detected and the requested interaction lengths $\lambda\sim
  10$ kpc are very hard to justify.

  By our fluctuative hypothesis, as we discussed above,
  $\lambda_g\sim 10$ kpc naturally emerges by taking into
  account stochastic fluctuations of the granular components of a
  galaxy. Using, Eq.(\ref{yukawa1}) where we assume
  $\lambda_1=\lambda_g$, the arguments in \cite{sanders} and in
  \cite{eckhardt} are easily recovered.

  Besides, the anomalous, long-range acceleration reported in
  \cite{anderson} immediately outside the Solar System, could be explained
  considering a Yukawa correction in the Newtonian potential related to
  a length as  $\lambda_s$ which
  can be considered as  the typical range of interaction of a star as the
  Sun (a system with gravitationally bound planet).

\section{\normalsize \bf Discussion and Conclusions}

In this paper, we have discussed the possibility that the
characteristic sizes of astrophysical self-gravitating systems
could be deduced by scaling laws relating the observed
macroscopic dimensions to the microscopic fundamental scales.
These scaling laws emerge taking into account the stochastic
behaviour of complex systems whose characteristic sizes come out
from statistical fluctuations. The net effects are Yukawa-like
corrections to the Newtonian potential which become relevant  in
the range $r\sim \lambda$ and, in general, modify the ''strength"
of gravitational coupling (e.g. $G_{lab}=G_{\infty}(1+\alpha)$).
This fact could be connected to the well-known absence of
screening mechanisms for gravity (see, for example
\cite{bertolami} and reference therein,  for the discussion of
quantum gravity effects on large scale structures).

Before drawing the conclusions, we have to discuss the scales of
action involved. Another link between the quantum unit of action
$h$ and the cosmological scales is provided by the so--called
Eddington--Weinberg relation $h \cong G^{1/2}m^{3/2}R^{1/2}$,
where $G$ is the Newton gravitational constant, $m$ is the mass
of the nucleon, and $R$ is the radius of the Universe.
 If one takes for $R$ the various definitions of cosmological radius (Hubble
radius, causal radius, or last scattering radius)\cite{peebles}
which range from $R=10^{26} cm$ to $R=10^{30} cm$, one obtains a
value for the unit of action $\alpha$ ranging from $10^{-26}$ erg
sec to $10^{-27}$ erg sec which is usually  assumed to coincide
with the Planck constant $h$ in order of magnitude \cite{barrow}.

A similar relation can be deduced also for the self-gravitating
structures which we have discussed, if one inserts in the
equations for $A$ and $\alpha$ the gravitational energy $U(R)$
and a characteristic gravitational time ${\cal T}$ needed for the
relaxation of the system.
 In this case, a sort of
Eddington--Weinberg relation can be derived and appears to hold
also for galaxies, and other large scale structures as clusters
and super-clusters of galaxies yielding a microscopic unit of
action of the order of $10^{-27\div -28}$ erg sec. However, in
the framework of our model, this quasi--coincidence is of no real
significance, because, as explained above, the correct way to
compute the characteristic energy, time, action, and length
scales for the astrophysical structures  must depend only on
observational kinetic and photometric quantities, and not on
energy and time scales explicitly dependent on the characteristic
dimension $R$ that one is seeking to predict. Therefore, the
fundamental relations in our model, are the set of
Eqs.(\ref{1})-(\ref{4}) and (\ref{10}) obtained above, and not
the Eddington--Weinberg relation. This is  more clear if one
moves to consider a planetary system or a globular cluster which
are not very large astrophysical structures. As discussed above,
the scheme presented in this paper $(R=\lambda_{c}\sqrt{N})$
perfectly applies also to these cases, yielding the correct
values for the characteristic radii. However, if one tries to
reinterpret them in terms of the Eddington--Weinberg relation,
one finds a microscopic unit of action of the order of $10^{-32}$
erg sec, thus devoid of any physical significance (according to
our present knowledge of the microscopic world). Therefore, the
numerical coincidence up to two orders of magnitude of the
Eddington--Weinberg relations for the Universe and for large
scale structures is, in our opinion, purely accidental.
Therefore, as we have shown, what is really significant in our
model, both on astrophysical and cosmological scales, are the
micro/macro scaling relations and connectivity factors, that is
Eqs. (\ref{2}), (\ref{4}), and (\ref{10}), which hold true for
all systems and are built starting from the basic  assumption of
nuclear granularity, from the statistical fluctuative hypothesis
(which provides the factor $\sqrt{N}$), and from purely kinetic,
statistical and photometric observational quantities.

 In conclusion, the typical hierarchical  sizes of astrophysical structures
  could be explained by taking into account a fluctuative
  hypothesis which yields typical interaction ranges for the given
  granular components. Dynamics is implemented by a non-Newtonian
  gravitational potential where Yukawa corrections effectively act
  at that typical scale. However, as sketched in Sec.2,
  the value of the gravitational coupling is
  different at the various distances depending on the interaction
  ranges $\lambda_c, \lambda_s,\lambda_g$. Finally, we want to
  stress that no vector boson or additional particle have been
  introduced and, thanks to the fluctuative hypothesis, the standard
  nucleons can completely account for sizes and stability of
  astrophysical structures.
  The results are in agreement with the
statistical approach to the structure of spacetime (the so called
``statistical geometry" as it is widely discussed in \cite{roy}).

\vspace{2. cm}

\begin{centerline}
{\bf REFERENCES}
\end{centerline}

\vfill
\end{document}